# Numerical modelling of microchannel gas flows in the transition flow regime using the cascaded lattice Boltzmann method

Qing Liu and Ya-Ling He

*Key Laboratory of Thermo-Fluid Science and Engineering of Ministry of Education, School of Energy and Power Engineering, Xi'an Jiaotong University, Xi'an, Shaanxi, 710049, China*
(Qing Liu: qingliu1983@stu.xjtu.edu.cn; Ya-Ling He: yalinghe@mail.xjtu.edu.cn)

**Abstract**

**Purpose –** The purpose of this paper is to present a cascaded lattice Boltzmann (CLB) method with the Bosanquet-type effective viscosity for simulating two-dimensional (2D) microchannel gas flows in the transition flow regime.

**Design/methodology/approach –** In the CLB method, a Bosanquet-type effective viscosity is employed to account for the rarefaction effect on the gas viscosity. To gain accurate simulations and to match the Bosanquet-type effective viscosity, the combined bounce-back/specular-reflection boundary scheme with a modified second-order slip boundary condition is adopted in the CLB method. The present method is applied to study gas flow in a microchannel with periodic boundary condition and gas flow in a long microchannel (the ratio of the length to the height $L/H = 100$) with pressure boundary condition over a wide range of Knudsen numbers. The influence of the rarefaction effect on the velocity profile, the mass flow rate, and the non-linear pressure deviation distribution along the long microchannel are investigated.

**Findings –** The predicted results, including the velocity profile, the mass flow rate, and the non-linear pressure deviation distribution along the long microchannel, agree well with the solution of the linearized Boltzmann equation, the DSMC and IP-DSMC results, and the experimental data reported in the literature. The characteristic flow behaviors of microchannel gas flows, such as the Knudsen minimum phenomenon (captured by the CLB method at $Kn \approx 0.9$), and the distributions of the streamwise and spanwise velocities, are well captured.

**Originality/value –** A new lattice Boltzmann method is developed for simulating microchannel gas flows in the transition flow regime. The present method can be served as a numerically accurate and computationally efficient numerical tool for simulating microscale rarefied gas flows.

**Keywords** – Lattice Boltzmann method, Cascaded collision operator, Microchannel gas flows, Transition flow regime, Bosanquet-type effective viscosity

## 1. Introduction

Understanding the fundamental flow characteristics of microscale rarefied gas flows is of crucial importance in designing, optimizing, fabricating and operating advanced microfluidic devices in micro-electro-mechanical systems (MEMS). Over the last few decades, microscale rarefied gas flows have attracted considerable research attention owing to the rapid progress of fabrication techniques in MEMS devices (e.g., microchannels, micropipes, microturbines, and microbearings) (Ho and Tai, 1998; Karniadakis *et al*, 2006; Barber and Emerson, 2006; Zhang *et al*., 2012). Typically, gas flows in microfluidic devices can be characterized by the Knudsen number $Kn = \lambda/H$ (the ratio of the mean free path $\lambda$ of the gas molecules to the characteristic length $H$ of the flow system). The Knudsen number serves as a criterion in indicating the degree of the rarefaction effects of gas flows. According to the magnitude of $Kn$, gas flows can be empirically classified into four regimes (Tsien, 1946; Gad-el-Hak, 1999; Wang *et al*., 2016): the continuum flow regime ($Kn < 0.001$), the slip flow regime ($0.001 < Kn < 0.1$), the transition flow regime ($0.1 < Kn < 10$), and the free molecular flow regime ($Kn > 10$). It is widely accepted that the Boltzmann equation (BE) can be used to describe gas flows ranging from the continuum flow regime to the free molecular flow regime. In the literature, it is well understood that the continuum-based Navier-Stokes (N-S) equations in conjunction with slip boundary conditions remain valid up to $Kn = 0.1$ or thereabouts (Gad-el-Hak, 1999; Aubert and Colin, 2001). However, for $Kn > 0.1$ (the transition and free molecular flows), the flow characteristics are dominated by the rarefaction effects and the classical descriptions based on the traditional N-S equations are no longer valid because of the breakdown of the continuum and thermodynamic equilibrium hypotheses (Barber and Emerson, 2006; Li *et al*., 2011; Zhang *et al*., 2012), and therefore, the Boltzmann equation (BE) must be considered to analyze such flows (Cercignani, 1990; Cercignani *et al*., 2004).

For gas flows in MEMS devices where the geometric size of the flow domain is small, the Knudsen number $Kn$ is relatively large, and such flows usually fall into the slip and transition flow regimes (Schaaf and Chambré, 1961). Due to technical advances, gas flows in microfluidic systems have been studied experimentally by many researchers in the past several decades (Arkilic *et al*., 1997; Maurer *et al*., 2003; Colin *et al*., 2004; Colin, 2005). In addition to the experimental investigations, theoretical and numerical analyses usually play an important role in revealing underlying mechanisms and providing design guidelines for practical engineering applications in MEMS, and even interpreting experimental observed phenomena. As reported by Cercignani (1990), the BE is applicable for gas flows ranging from the continuum flow regime to the free molecular flow regime. Theoretically, rarefied gas flows in the slip flow and transition flow regimes can be described by directly solving the BE or simulated by using the direct simulation Monte Carlo (DSMC) method (Bird, 1994). However, it has been demonstrated that the solution of the BE is impractical except for a few cases, while the DSMC method usually suffers from highly statistical noise and extremely expensive computational cost in solving practical engineering problems. Therefore, many efforts have been made to develop numerically accurate and computationally efficient numerical methods based on the BE in the kinetic theory. Since the late 1990s, a variety of BE-based mesoscopic numerical methods, such as the discrete-velocity method (DVM) (Sharipov, 1999; Naris *et al*., 2004), the lattice Boltzmann (LB) method (Nie *et al*., 2002; Lim *et al*., 2002), the gas-kinetic scheme (GKS) (Xu and Li, 2004), and the discrete unified gas kinetic scheme (DUGKS) (Guo *et al*., 2013), have been developed to study microscale rarefied gas flows. Among these BE-based mesoscopic numerical methods, the LB method has attracted significant attention in studying microscale rarefied gas flows since 2002 (Nie *et al*., 2002; Lim *et al*., 2002; Succi, 2002; Guo *et al*., 2006, 2008; Tang *et al*., 2007; Verhaeghe *et al*., 2009; Zhang, 2011; Li *et al*., 2011; Zhuo and Zhong, 2013; Bo *et al*., 2015).

Historically, the LB method (McNamara and Zanetti, 1988; Higuera *et al*., 1989; Chen *et al*., 1991; Qian *et al*., 1992) originated from the lattice gas automata (LGA) method (Frisch *et al*., 1986), which can be viewed as a simplified, fictitious version of the molecular dynamics (MD) model in which the time, space, and particle velocities are all discrete. Later He and Luo (1997) proved that the LB equation can be rigorously obtained from the BE in the kinetic theory. Owning to its kinetic nature, the LB method possesses many distinctive advantages over the traditional numerical methods (Succi, 2008), and it has gained great success in simulating complex fluid flows and modeling complex physics in fluids (Benzi *et al*., 1992; Chen and Doolen, 1998; Mohamad, 2011; Ali Rabienataj Darzi *et al*., 2013; Succi, 2015; Sun *et al*., 2015; Gao *et al*., 2015; Liu *et al*., 2016; Li *et al*., 2016). In recent years, the cascaded lattice Boltzmann (CLB) method (Geier *et al*., 2006; Premnath and Banerjee, 2009; Geller *et al*., 2013; Lycett-Brown and Luo, 2014; Ning *et al*., 2015) has also attracted much attention. The CLB method was first proposed by Geier *et al*. (2006). The key point in this method is the use of cascaded collision operator, in which the collision process is performed in terms of central moments (moments obtained by shifting the particle velocity by the local macroscopic fluid velocity) in an ascending order by order at different relaxation rates. Galilean invariance can be naturally prescribed in the CLB method. Numerical experiments (Ning *et al*., 2015) demonstrate that the cascaded collision model results in significant stability improvements when compared with the BGK collision model (Qian *et al*., 1992) and MRT collision model (Lallemand and Luo, 2000).

In the past decade, the LB method has attracted significant attention in studying microscale rarefied gas flows, and it has been well accepted as an efficient and useful numerical approach to simulate gaseous flows in various microfluidic systems (Zhang, 2011). As mentioned before, in the CLB method, Galilean invariance is naturally prescribed and different central moments can be relaxed at different rates, thus it has freedom to adjust higher-order discretization errors due to boundary conditions. It is expected that the CLB method can be used to simulate microscale rarefied gas flows in the transition flow regime. To the best of our knowledge, no studies have been reported in the open literature on microscale rarefied gas flows by using the CLB method. Hence, the purpose of this paper is to present a CLB method with the Bosanquet-type effective viscosity for simulating microchannel gas flows in the transition flow regime. In order to gain accurate simulations and match the Bosanquet-type effective viscosity, the combined bounce-back/specular-reflection (CBBSR) scheme with a modified second-order slip boundary condition is employed in the CLB method.

The rest of this paper is organized as follows. In Section 2, the CLB method for simulating microscale rarefied gas flows is presented in detail. Numerical simulations of gas flow in a microchannel with periodic boundary condition and gas flow in a long microchannel with pressure boundary condition are carried out in Section 3. Finally, some conclusions are made in Section 4.

## 2. The CLB method for microscale rarefied gas flows

### 2.1 The CLB model

In this subsection, the CLB model with forcing term (Premnath and Banerjee, 2009) will be briefly introduced. For two-dimensional (2D) microscale rarefied gas flows considered in this work, the two-dimensional nine-velocity (D2Q9) lattice model is employed. The nine discrete velocities $\{\mathbf{e}_i | i = 0, 1, \ldots, 8\}$ of the D2Q9 lattice are given by (Qian et al., 1992)

$$\mathbf{e}_i = \begin{cases} (0,0), & i = 0, \\ \left(\cos\left[(i-1)\pi/2\right], \sin\left[(i-1)\pi/2\right]\right)c, & i = 1-4, \\ \left(\cos\left[(2i-9)\pi/4\right], \sin\left[(2i-9)\pi/4\right]\right)\sqrt{2}c, & i = 5-8, \end{cases} \quad (1)$$

where $c = \delta_x/\delta_t$ is the lattice speed with $\delta_t$ and $\delta_x$ being the discrete time step and lattice spacing, respectively. The lattice speed $c$ is set to be 1 ($\delta_x = \delta_t$) in this work.

In general, the CLB equation with a semi-implicit treatment of the forcing term can be written as

$$f_i\left(\mathbf{x} + \mathbf{e}_i\delta_t, t + \delta_t\right) = f_i\left(\mathbf{x}, t\right) + \Omega_i^C\Big|_{(\mathbf{x},t)} + \frac{\delta_t}{2}\left[S_i\Big|_{(\mathbf{x},t)} + S_i\Big|_{(\mathbf{x}+\mathbf{e}_i\delta_t, t+\delta_t)}\right], \quad (2)$$

where $f_i$ is the discrete density distribution function, $S_i$ is the forcing term representing the effect of an external force field $\mathbf{F} = (F_x, F_y)$, and $\Omega_i^C$ is the collision term. In cascaded collision model, the collision term $\Omega_i^C$ can be expressed as $\Omega_i^C \equiv \Omega_i^C(\mathbf{f}, \hat{\mathbf{g}}) = (\mathbf{K} \cdot \hat{\mathbf{g}})_i$, in which $\mathbf{f} = |f\rangle = (f_0, f_1, \ldots, f_8)^T$ is the vector of the discrete density distribution functions, $\hat{\mathbf{g}} = |\hat{g}\rangle = (\hat{g}_0, \hat{g}_1, \ldots, \hat{g}_8)^T$ is the vector of the unknown collision kernels to be determined later, and $\mathbf{K}$ is an orthogonal matrix given by

$$\mathbf{K} = \left[|1\rangle, |e_x\rangle, |e_y\rangle, 3|e_x^2 + e_y^2\rangle - 4|1\rangle, |e_x^2 - e_y^2\rangle, |e_x e_y\rangle, -3|e_x^2 e_y\rangle + 2|e_y\rangle, \right.$$
$$\left. -3|e_x e_y^2\rangle + 2|e_x\rangle, 9|e_x^2 e_y^2\rangle - 6|e_x^2 + e_y^2\rangle + 4|1\rangle\right]$$

$$= \begin{bmatrix} 1 & 0 & 0 & -4 & 0 & 0 & 0 & 0 & 4 \\ 1 & 1 & 0 & -1 & 1 & 0 & 0 & 2 & -2 \\ 1 & 0 & 1 & -1 & -1 & 0 & 2 & 0 & -2 \\ 1 & -1 & 0 & -1 & 1 & 0 & 0 & -2 & -2 \\ 1 & 0 & -1 & -1 & -1 & 0 & -2 & 0 & -2 \\ 1 & 1 & 1 & 2 & 0 & 1 & -1 & -1 & 1 \\ 1 & -1 & 1 & 2 & 0 & -1 & -1 & 1 & 1 \\ 1 & -1 & -1 & 2 & 0 & 1 & 1 & 1 & 1 \\ 1 & 1 & -1 & 2 & 0 & -1 & 1 & -1 & 1 \end{bmatrix}. \quad (3)$$

For the sake of brevity, the Dirac notations $\langle \cdot |$ and $| \cdot \rangle$ are used to denote the row and column vectors, respectively. Eq. (2) is implicit and cannot be directly implemented in simulations. By introducing a transformed distribution function $\bar{f}_i = f_i - 0.5\delta_t S_i$, the implicitness of the CLB equation (2) can be eliminated, which yields

$$\tilde{\bar{f}}_i(\mathbf{x}, t) = \bar{f}_i(\mathbf{x}, t) + \Omega_i^C\Big|_{(\mathbf{x},t)} + \delta_t S_i\Big|_{(\mathbf{x},t)}, \quad (4)$$

$$\bar{f}_i\left(\mathbf{x}+\mathbf{e}_i\delta_t, t+\delta_t\right) = \tilde{\bar{f}}_i\left(\mathbf{x}, t\right), \tag{5}$$

where Eqs. (4) and (5) denote the collision step and streaming step, respectively, and $\tilde{\bar{f}}_i$ is the post-collision distribution function. According to the orthogonal matrix $\mathbf{K}$, the collision step (4) can be expanded as follows:

$$\tilde{\bar{f}}_0 = \bar{f}_0 + \left[\hat{g}_0 - 4(\hat{g}_3 - \hat{g}_8)\right] + \delta_t S_0,$$

$$\tilde{\bar{f}}_1 = \bar{f}_1 + \left[\hat{g}_0 + \hat{g}_1 - \hat{g}_3 + \hat{g}_4 + 2(\hat{g}_7 - \hat{g}_8)\right] + \delta_t S_1,$$

$$\tilde{\bar{f}}_2 = \bar{f}_2 + \left[\hat{g}_0 + \hat{g}_2 - \hat{g}_3 - \hat{g}_4 + 2(\hat{g}_6 - \hat{g}_8)\right] + \delta_t S_2,$$

$$\tilde{\bar{f}}_3 = \bar{f}_3 + \left[\hat{g}_0 - \hat{g}_1 - \hat{g}_3 + \hat{g}_4 - 2(\hat{g}_7 + \hat{g}_8)\right] + \delta_t S_3,$$

$$\tilde{\bar{f}}_4 = \bar{f}_4 + \left[\hat{g}_0 - \hat{g}_2 - \hat{g}_3 - \hat{g}_4 - 2(\hat{g}_6 + \hat{g}_8)\right] + \delta_t S_4, \tag{6}$$

$$\tilde{\bar{f}}_5 = \bar{f}_5 + \left[\hat{g}_0 + \hat{g}_1 + \hat{g}_2 + 2\hat{g}_3 + \hat{g}_5 - \hat{g}_6 - \hat{g}_7 + \hat{g}_8\right] + \delta_t S_5,$$

$$\tilde{\bar{f}}_6 = \bar{f}_6 + \left[\hat{g}_0 - \hat{g}_1 + \hat{g}_2 + 2\hat{g}_3 - \hat{g}_5 - \hat{g}_6 + \hat{g}_7 + \hat{g}_8\right] + \delta_t S_6,$$

$$\tilde{\bar{f}}_7 = \bar{f}_7 + \left[\hat{g}_0 - \hat{g}_1 - \hat{g}_2 + 2\hat{g}_3 + \hat{g}_5 + \hat{g}_6 + \hat{g}_7 + \hat{g}_8\right] + \delta_t S_7,$$

$$\tilde{\bar{f}}_8 = \bar{f}_8 + \left[\hat{g}_0 + \hat{g}_1 - \hat{g}_2 + 2\hat{g}_3 - \hat{g}_5 + \hat{g}_6 - \hat{g}_7 + \hat{g}_8\right] + \delta_t S_8.$$

Considering the effect of the external force, the collision kernels $\{\hat{g}_i \mid i=0, 1, \ldots, 8\}$ can be obtained as follows

$$\hat{g}_0 = \hat{g}_1 = \hat{g}_2 = 0,$$

$$\hat{g}_3 = \frac{s_3}{12}\left\{\frac{2}{3}\rho + \rho\left(u_x^2 + u_y^2\right) - \left(\hat{\kappa}'_{xx} + \hat{\kappa}'_{yy}\right) - \frac{1}{2}\rho\left(2F_x u_x + 2F_y u_y\right)\right\},$$

$$\hat{g}_4 = \frac{s_4}{4}\left\{\rho\left(u_x^2 - u_y^2\right) - \left(\hat{\kappa}'_{xx} - \hat{\kappa}'_{yy}\right) - \frac{1}{2}\rho\left(2F_x u_x - 2F_y u_y\right)\right\},$$

$$\hat{g}_5 = \frac{s_5}{4}\left\{\rho u_x u_y - \hat{\kappa}'_{xy} - \frac{1}{2}\rho\left(F_x u_y + F_y u_x\right)\right\},$$

$$\hat{g}_6 = \frac{s_6}{4}\left\{2\rho u_x^2 u_y + \hat{\kappa}'_{xxy} - 2u_x \hat{\kappa}'_{xy} - u_y \hat{\kappa}'_{xx} - \frac{1}{2}\rho\left(F_y u_x^2 + 2F_x u_x u_y\right)\right\}$$

$$\quad - \frac{1}{2}u_y\left(3\hat{g}_3 + \hat{g}_4\right) - 2u_x \hat{g}_5, \tag{7}$$

$$\hat{g}_7 = \frac{s_7}{4}\left\{2\rho u_x u_y^2 + \hat{\kappa}'_{xyy} - 2u_y \hat{\kappa}'_{xy} - u_x \hat{\kappa}'_{yy} - \frac{1}{2}\rho\left(F_x u_y^2 + 2F_y u_y u_x\right)\right\}$$

$$\quad - \frac{1}{2}u_x\left(3\hat{g}_3 - \hat{g}_4\right) - 2u_y \hat{g}_5,$$

$$\hat{g}_8 = \frac{s_8}{4}\left\{\frac{1}{9}\rho + 3\rho u_x^2 u_y^2 - \left[\hat{\tilde{\kappa}}'_{xxyy} - 2u_x\hat{\tilde{\kappa}}'_{xyy} - 2u_y\hat{\tilde{\kappa}}'_{xxy} + u_x^2\hat{\tilde{\kappa}}'_{yy} + u_y^2\hat{\tilde{\kappa}}'_{xx} + 4u_x u_y\hat{\tilde{\kappa}}'_{xy}\right]\right.$$

$$\left. -\frac{1}{2}\rho\left(2F_x u_x u_y^2 + 2F_y u_y u_x^2\right)\right\} - 2\hat{g}_3 - \frac{1}{2}u_y^2\left(3\hat{g}_3 + \hat{g}_4\right) - \frac{1}{2}u_x^2\left(3\hat{g}_3 - \hat{g}_4\right)$$

$$-4u_x u_y \hat{g}_5 - 2u_y \hat{g}_6 - 2u_x \hat{g}_7,$$

where $\{s_i | i=3,4,\ldots,8\}$ are relaxation rates, and $\hat{\tilde{\kappa}}'_{x^m y^n} = \langle e_x^m e_y^n | \bar{f}\rangle$ ($m,n \in \{0,1,2\}$, and $\langle e_x^m e_y^n | \bar{f}\rangle$ denotes the inner product $\sum_{i=0}^{8} e_{ix}^m e_{iy}^n \bar{f}_i$) is the discrete raw moment of the transformed distribution functions of order $(m+n)$. The discrete central moment of the transformed distribution functions of order $(m+n)$ is defined by $\hat{\tilde{\kappa}}_{x^m y^n} = \langle (e_x - u_x)^m (e_y - u_y)^n | \bar{f}\rangle$ (Geier *et al.*, 2006; Premnath and Banerjee, 2009). According to the binomial theorem, central moment of a given order is algebraic combination of raw moments of different orders, with the highest order being equal to that of the central moment. In computations, the collision step of the CLB method is actually performed in terms of the raw moments. Due to the cascaded nature of the central moment approach, the collision kernel $\hat{g}_i$ satisfies $\hat{g}_i \equiv \hat{g}_i(\mathbf{f}, \hat{g}_\beta)$, $\beta = 0,1,\ldots,i-1$. For the D2Q9 model, the discrete raw moments $\hat{\tilde{\kappa}}'_{x^m y^n}$ can be expressed as follows:

$$\hat{\tilde{\kappa}}'_0 = \langle 1 | \bar{f}\rangle = \rho,$$

$$\hat{\tilde{\kappa}}'_x = \langle e_x | \bar{f}\rangle = \rho u_x - \frac{1}{2}\rho F_x,$$

$$\hat{\tilde{\kappa}}'_y = \langle e_y | \bar{f}\rangle = \rho u_y - \frac{1}{2}\rho F_y,$$

$$\hat{\tilde{\kappa}}'_{xx} = \langle e_x^2 | \bar{f}\rangle = \sum_i^{\{1,3,5,6,7,8\}} \bar{f}_i,$$

$$\hat{\tilde{\kappa}}'_{yy} = \langle e_y^2 | \bar{f}\rangle = \sum_i^{\{2,4,5,6,7,8\}} \bar{f}_i, \qquad (8)$$

$$\hat{\tilde{\kappa}}'_{xy} = \langle e_x e_y | \bar{f}\rangle = \sum_i^{\{5,7\}} \bar{f}_i - \sum_i^{\{6,8\}} \bar{f}_i,$$

$$\hat{\tilde{\kappa}}'_{xxy} = \langle e_x^2 e_y | \bar{f}\rangle = \sum_i^{\{5,6\}} \bar{f}_i - \sum_i^{\{7,8\}} \bar{f}_i,$$

$$\hat{\tilde{\kappa}}'_{xyy} = \langle e_x e_y^2 | \bar{f}\rangle = \sum_i^{\{5,8\}} \bar{f}_i - \sum_i^{\{6,7\}} \bar{f}_i,$$

$$\hat{\tilde{\kappa}}'_{xxyy} = \langle e_x^2 e_y^2 | \bar{f}\rangle = \sum_i^{\{5,6,7,8\}} \bar{f}_i.$$

The forcing term $\mathbf{S} = |S\rangle$ can be obtained via $\mathbf{S} = \mathbf{T}^{-1}\hat{\mathbf{S}}$, in which $\hat{\mathbf{S}} = |\hat{S}\rangle$ is given by

$$\hat{\mathbf{S}} = \begin{bmatrix} 0 \\ \rho F_x \\ \rho F_y \\ 2\rho(u_x F_x + u_y F_y) \\ 2\rho(u_x F_x - u_y F_y) \\ \rho(u_x F_y + u_y F_x) \\ \rho F_y u_x^2 + 2\rho F_x u_x u_y \\ \rho F_x u_y^2 + 2\rho F_y u_y u_x \\ 2\rho F_x u_x u_y^2 + 2\rho F_y u_y u_x^2 \end{bmatrix}. \tag{9}$$

$\mathbf{T}$ is a non-orthogonal transformation matrix defined by

$$\mathbf{T} = \left[ |1\rangle, |e_x\rangle, |e_y\rangle, |e_x^2 + e_y^2\rangle, |e_x^2 - e_y^2\rangle, |e_x e_y\rangle, |e_x^2 e_y\rangle, |e_x e_y^2\rangle, |e_x^2 e_y^2\rangle \right]^{\mathrm{T}}.$$

$$= \begin{bmatrix} 1 & 1 & 1 & 1 & 1 & 1 & 1 & 1 & 1 \\ 0 & 1 & 0 & -1 & 0 & 1 & -1 & -1 & 1 \\ 0 & 0 & 1 & 0 & -1 & 1 & 1 & -1 & -1 \\ 0 & 1 & 1 & 1 & 1 & 2 & 2 & 2 & 2 \\ 0 & 1 & -1 & 1 & -1 & 0 & 0 & 0 & 0 \\ 0 & 0 & 0 & 0 & 0 & 1 & -1 & 1 & -1 \\ 0 & 0 & 0 & 0 & 0 & 1 & 1 & -1 & -1 \\ 0 & 0 & 0 & 0 & 0 & 1 & -1 & -1 & 1 \\ 0 & 0 & 0 & 0 & 0 & 1 & 1 & 1 & 1 \end{bmatrix}. \tag{10}$$

The equilibrium distribution function $f_i^{eq}$ can be obtained via $\mathbf{f}^{eq} = \mathbf{T}^{-1}\hat{\mathbf{f}}^{eq}$, in which $\hat{\mathbf{f}}^{eq} = \left| \hat{f}^{eq} \right\rangle$ is given by

$$\hat{\mathbf{f}}^{eq} = \begin{bmatrix} \rho \\ \rho u_x \\ \rho u_y \\ \frac{2}{3}\rho + \rho(u_x^2 + u_y^2) \\ \rho(u_x^2 - u_y^2) \\ \rho u_x u_y \\ \frac{1}{3}\rho u_y + \rho u_x^2 u_y \\ \frac{1}{3}\rho u_x + \rho u_x u_y^2 \\ \frac{1}{9}\rho + \frac{1}{3}\rho(u_x^2 + u_y^2) + \rho u_x^2 u_y^2 \end{bmatrix}. \tag{11}$$

The macroscopic fluid density $\rho$ and velocity $\mathbf{u}$ can be calculated by

$$\rho = \sum_{i=0}^{8} f_i = \sum_{i=0}^{8} \bar{f}_i, \tag{12}$$

$$\rho \mathbf{u} = \sum_{i=0}^{8} \mathbf{e}_i f_i = \sum_{i=0}^{8} \mathbf{e}_i \bar{f}_i + \frac{\delta_t}{2}\rho \mathbf{F}. \tag{13}$$

The pressure $p$ is defined by $p = \rho RT = \rho c_s^2$, where $c_s = \sqrt{RT} = c/\sqrt{3}$ is the lattice sound speed, in which $R$ is the gas constant and $T$ is the temperature. The kinematic viscosity $\upsilon$ and bulk

viscosity $\zeta$ are given by

$$\upsilon = c_s^2 \left(\frac{1}{s_\upsilon} - \frac{1}{2}\right)\delta_t, \quad \zeta = c_s^2 \left(\frac{1}{s_b} - \frac{1}{2}\right)\delta_t, \quad (14)$$

respectively, where $s_4 = s_5 = s_\upsilon$, and $s_3 = s_b$. The dynamic viscosity $\mu$ is given by $\mu = \rho\upsilon$. The rest of the relaxation rates can be tuned independently to improve the numerical stability (Geier *et al.*, 2006; Premnath and Banerjee, 2009). Obviously, the CLB method has freedom to adjust higher-order discretization errors due to boundary conditions, which is important in simulating microscale rarefied gas flows. The cascaded collision term $\Omega_i^C$ is constructed in such a way that the central moments can be relaxed independently at different relaxation rates. From this point of view, the CLB method can be regarded as a central-moment-based MRT scheme, whereas by contrast the standard MRT method (Lallemand and Luo, 2000) is based on raw moments.

*2.2 Bosanquet-type effective viscosity*

For microscale gas flows, the Knudsen number is the most important characteristic parameter, and the relationship between the dynamic viscosity (or relaxation rate) and the Knudsen number must be given appropriately. In the kinetic theory, the relationship between the dynamic viscosity $\mu$ and the mean free path $\lambda$ can be expressed as (Cercignani, 1990)

$$\lambda = \frac{\mu}{p}\sqrt{\frac{\pi RT}{2}}. \quad (15)$$

As pointed out by Guo *et al.* (2006), the above relationship is only valid for rarefied gas flows in unbounded systems. For rarefied gas flows in bounded systems, the dynamic viscosity and mean free path given by Eq. (15) are questionable because the presence of walls will reduce the local mean free path at the near wall regions (Guo *et al.*, 2006, 2008; Li *et al.*, 2011). In order to reflect the effect of gas molecule/wall interactions, the Bosanquet-type effective viscosity (Beskok and Karniadakis, 1999; Michalis *et al.*, 2010) is introduced into the CLB method. The Bosanquet-type effective viscosity can be expressed as

$$\mu_e = \frac{\mu}{1 + aKn}, \quad (16)$$

where $a$ is the rarefaction factor. Accordingly, an effective mean free path $\lambda_e$ can be determined by $\lambda_e = (\mu_e/p)\sqrt{\pi RT/2}$. The rarefaction factor $a$ is not constant but depends on $Kn$. However, as reported by Michalis *et al.* (2010), this dependence is rather weak over the majority of the transition flow regime, suggesting an effective value close to 2. Based on Michalis *et al.*'s study, we use $\mu_e = \mu/(1+aKn)$ with $a = 2$ in the present study. According to Eqs. (15) and (16), the Bosanquet-type effective viscosity $\mu_e$ can be determined from the Knudsen number $Kn$ as

$$\mu_e = \frac{c}{3}\sqrt{\frac{6}{\pi}}\frac{\rho KnN\delta_x}{1+aKn}, \tag{17}$$

where $N = H/\delta_x$ is the lattice number in the direction of the characteristic length. According to Eqs. (14) and (17), the relaxation rate $s_\nu$ can be determined by

$$s_\nu^{-1} = \frac{1}{2} + \sqrt{\frac{6}{\pi}}\frac{NKn}{(1+aKn)}. \tag{18}$$

Through Eq. (18), the Bosanquet-type effective viscosity $\mu_e$ is produced in the CLB method.

*2.3 Boundary condition*

When the Bosanquet-type effective viscosity is used, the following modified second-order slip boundary condition (Li *et al.*, 2011) should be considered:

$$u_s = B_1\sigma_\nu\lambda_e\left.\frac{\partial u}{\partial \mathbf{n}}\right|_w - B_2\lambda_e^2\left.\frac{\partial^2 u}{\partial \mathbf{n}^2}\right|_w, \tag{19}$$

where $u_s$ is the slip velocity, $B_1$ and $B_2$ are the first-order and second-order slip coefficients, respectively, $\mathbf{n}$ is the unit vector normal to the wall, the subscript w represents the quantity at the wall, and $\sigma_\nu = (2-\sigma)/\sigma$, in which $\sigma$ is the TMAC (tangential momentum accommodation coefficient).

Usually, the macroscopic slip boundary condition cannot be directly implemented in the LB method. In the present CLB method, the CBBSR boundary scheme (Succi, 2002; Guo *et al.*, 2008; Li *et al.*, 2011) is adopted to realize the modified second-order slip boundary condition given by Eq. (19). Taking the treatment at the bottom wall for example, the unknown distribution functions ($\bar{f}_2$, $\bar{f}_5$, and $\bar{f}_6$) at $J = 1$ (the wall is plated at $J = 0.5$) are given by

$$\bar{f}_2 = \tilde{\bar{f}}_4, \quad \bar{f}_5 = r_b\tilde{\bar{f}}_7 + (1-r_b)\tilde{\bar{f}}_8, \quad \bar{f}_6 = r_b\tilde{\bar{f}}_8 + (1-r_b)\tilde{\bar{f}}_7, \tag{20}$$

where $\tilde{\bar{f}}_i$ ($i = 2,5,6$) are the post-collision distribution functions at $J = 1$, and $r_b \in [0,1]$ is the portion of the bounce-back part in the combination. To match the modified second-order slip boundary condition at the macroscopic level, according to Li *et al.* (2011), the parameter $r_b$ and the relaxation

rate $s_q$ ($s_6 = s_7 = s_q$) should be chosen as follows:

$$r_b = \frac{1}{1 + B_1 \sigma_v \sqrt{\pi/6}}, \quad s_q^{-1} = \frac{1}{2} + \frac{3 + 4\pi \tilde{\tau}_q^2 B_2}{16 \tilde{\tau}_q}, \tag{21}$$

where $\tilde{\tau}_q = s_v^{-1} - 0.5$, in which $s_v$ is given by Eq. (18). Finally, the CLB model together with Eqs. (18), (20) and (21) constitute the present CLB method for simulating microscale rarefied gas flows.

## 3. Numerical simulations

In this section, the proposed CLB method will be applied to study gas flow in a microchannel with periodic boundary condition and gas flow in a long microchannel with pressure boundary condition over a wide range of Knudsen numbers. In the following simulations, we use $\delta_x = \delta_y = \delta_t = 1$, $a = 2$, $B_1 = (1 - 0.1817\sigma)$, and $B_2 = 0.55$. The relaxation rates are set as follows: $s_3 = 1.1$, $s_8 = 1.2$, $s_{4,5} = s_v$ is given by Eq. (18), and $s_{6,7} = s_q$ is given by Eq. (21).

### *3.1 Gas flow in a microchannel with periodic boundary condition*

First we consider the 2D gas flow in a microchannel with periodic boundary condition driven by a constant force in the transition flow regime, which has been studied by many researchers (Ohwada *et al.*, 1989; Hadjiconstantinou, 2003; Guo *et al.*, 2008; Li *et al.*, 2011; Zhuo and Zhong, 2013). The periodic boundary condition is imposed at the inlet and outlet of the channel, and the CBBSR boundary scheme is applied at the bottom and top walls with $\sigma = 1$ (gas molecules are reflected diffusely on the bottom and top walls). All computations are carried out on a uniform lattice $N_x \times N_y = 50 \times 50$, and the driven force $F_x$ is set to be $10^{-4}$.

Figure 1 shows the dimensionless velocity profiles across the channel at $Kn = 2k/\sqrt{\pi}$ with $k$ ranging from 0.1 to 10. The dimensionless velocity $U$ is defined by $U = u_x/\bar{u}_x$, where $\bar{u}_x = (1/H) \int_0^H u_x \, dy$ is the average velocity over the cross section of the channel. The benchmark solutions of the linearized BE obtained by Ohwada *et al.* (1989), the solutions of the conventional N-S equations using a second-order slip boundary condition given by Hadjiconstantinou (2003) (N-S-H solutions), and the numerical results obtained by Guo *et al.* (2008) using the MRT-LB method, are presented in Figure 1 for comparison. It can be seen that the N-S-H solutions significantly deviate from the linearized BE solutions when $Kn \geq 0.2257$. The MRT-LB results obtained by Guo *et al.* (2008) using Stops' expression of effective viscosity show a visible discrepancy from the linearized BE as $Kn \geq 1.1284$. On the other hand, the numerical results of the present CLB method agree well with the linearized BE solutions from $Kn = 0.1128$ to $4.5135$. At large Knudsen numbers ($Kn = 6.7703$, $9.0270$, and $11.2838$), the present numerical results and the linearized BE solutions show only slight differences. For comparison, the results of the Filter-matrix LB model (Zhuo and Zhong, 2013) using Bosanquet-type effective viscosity at large Knudsen numbers are also presented in the figure. It is clearly seen that our results are in good agreement with the Filter-matrix LB results at $Kn = 6.7703$, $9.0270$, and $11.2838$. To be more informative, the values of the slip velocity ($U_s$) predicted by the CLB method are also presented in the figure, and it can be seen that the slip velocity increases as the Knudsen number increases.

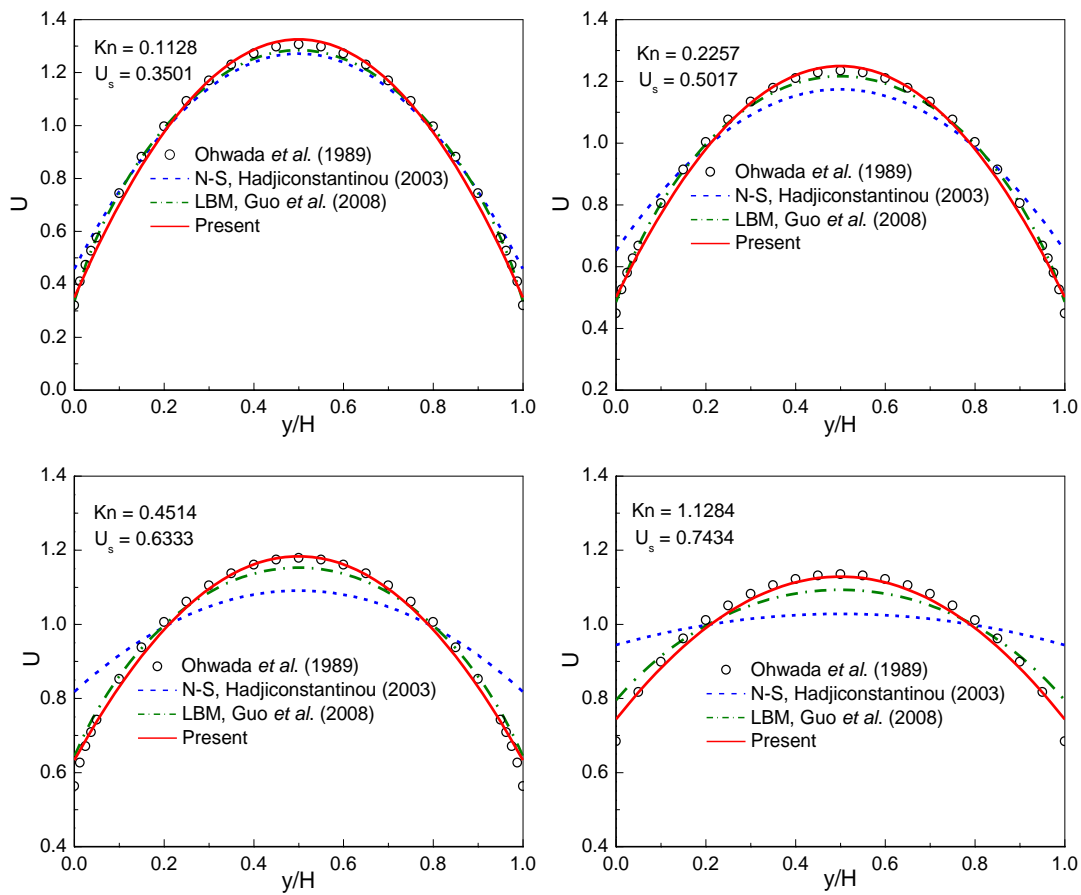

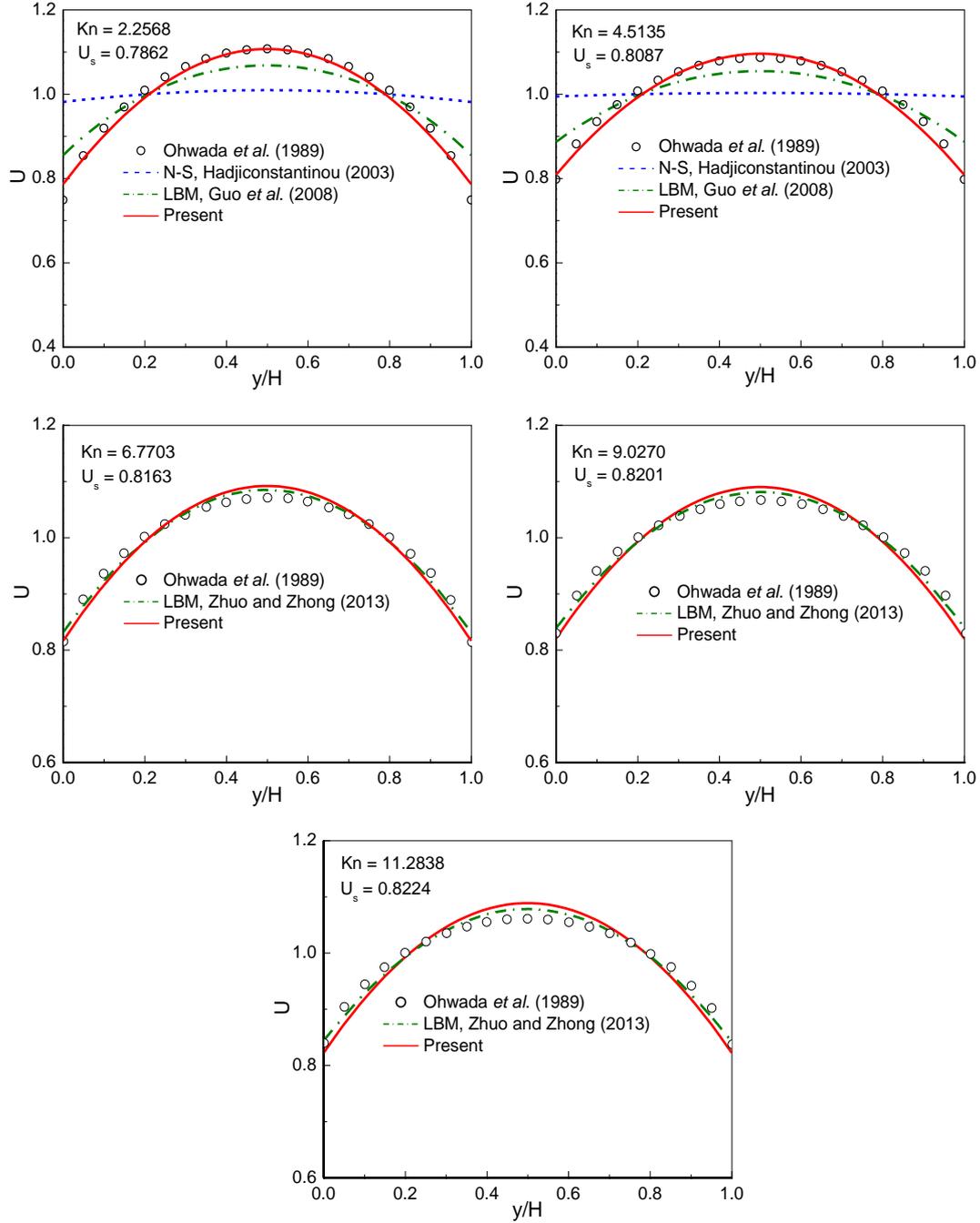

**Figure 1**. Dimensionless velocity profiles of gas flow in a microchannel with periodic boundary condition driven by a constant force ($\sigma = 1$).

To further validate the CLB method, in Figure 2 we plot the dimensionless flow rate $Q = \left(\int_0^H u_x \, dy\right) \Big/ \left(F_x H^2 \sqrt{RT/2}\big/p\right)$ (Hadjiconstantinou, 2003) against the Knudsen number. The linearized BE solutions given by Cercignani et al. (2004) using a variational approach, the N-S-H solutions given by Hadjiconstantinou (2003), and the MRT-LB results obtained by Guo et al. (2008), are presented in the figure for comparison. As can be seen, in comparison with the linearized BE solutions of Cercignani *et al.*, Hadjiconstantinou's approach can only predict an accurate flow rate for $Kn \leq 0.3$, whereas the CLB method can provide reasonable results up to $Kn \approx 5$. In addition, as reported by Karniadakis *et al.* (2006), there exists a minimum in the flow rate of microchannel flows at about $Kn \approx 1$. Cercignani *et al.*'s solution of the BE indicates that the Knudsen minimum phenomenon occurs at $Kn \approx 0.8$. In the present study, such a phenomenon is captured by the CLB method at $Kn \approx 0.9$.

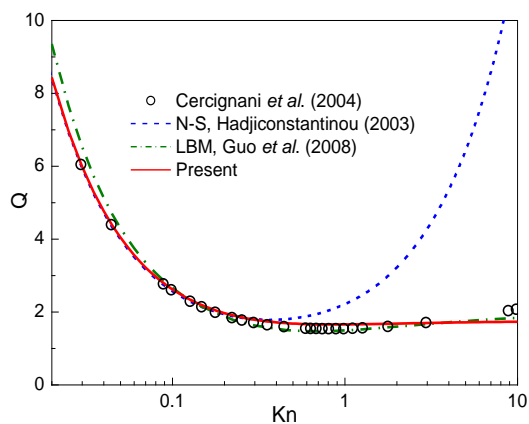

**Figure 2**. Dimensionless flow rate of gas flow in a microchannel with periodic boundary condition driven by a constant force ($\sigma = 1$).

### *3.2 Gas flow in a long microchannel with pressure boundary condition*

In this subsection, the CLB method is applied to study gas flow in a long microchannel with pressure boundary condition, which has become an important test case due to its wide engineering applications in MEMS (Dongari *et al.*, 2009). In this problem, a 2D microchannel with height $H$ and length $L$ is considered. The pressures at the inlet and outlet of the channel are $p_{in}$ and $p_{out}$, respectively, and the flow is driven by the substantial pressure drops. Following the literature (Shen *et al.*, 2004), the ratio of the length to the height $L/H$ is taken to be $100$. The local Knudsen number $Kn$ is determined by $Kn = Kn_{out} p_{out} / p(x)$, where $p(x)$ is the local pressure along the channel centerline, and $Kn_{out}$ is the Knudsen number at the outlet.

In our computations, a uniform lattice $N_x \times N_y = 2000 \times 20$ is employed. The CBBSR boundary scheme is applied to the bottom and top walls of the channel, and the inlet and outlet pressure boundary conditions are realized by using the consistent linear extrapolation scheme developed by Verhaeghe *et al.* (2009). The details of the consistent linear extrapolation scheme can be found in the literature (Verhaeghe *et al.*, 2009). We first consider the following three cases with $\sigma = 1$: (i) $Kn_{out} = 0.0194$,

$p_{in}/p_{out} = 1.4$; (ii) $Kn_{out} = 0.194$, $p_{in}/p_{out} = 2$; (iii) $Kn_{out} = 0.388$, $p_{in}/p_{out} = 2$. In Figures 3, 4 and 5, the dimensionless streamwise velocity $u_x/u_{x,\max}$ at the outlet and the pressure deviation $\delta p = (p - p_l)/p_{out}$ along the channel centerline are presented. Here, $u_{x,\max}$ is the maximum streamwise velocity at the outlet, and $p_l$ is the linear distributed pressure along the channel centerline defined by $p_l = p_{in} + (p_{in} - p_{out})x/L$. The analytical solutions (Arkilic et al., 1997) obtained from the N-S equations with first-order slip boundary condition (slip N-S solutions) and the DSMC and information-preservation DSMC (IP-DSMC) results of Shen et al. (2004) are also presented in the figures for comparison. For $Kn_{out} = 0.0194$ and $p_{in}/p_{out} = 1.4$ (see Figure 3), the flow is in the slip flow regime, and the profiles of the velocity and pressure deviation obtained by the present CLB method agree well with the DSMC, IP-DSMC, and slip N-S solutions. When $Kn_{out}$ increases to 0.194 with $p_{in}/p_{out} = 2.0$ (see Figure 4), the flow falls into the slip flow regime. It can be observed that the CLB results are in good agreement with the results of the DSMC and IP-DSMC methods. However, for the pressure deviation profile, the slip N-S solutions obviously deviate from those of the DSMC and IP-DSMC methods. For $Kn_{out} = 0.388$ and $p_{in}/p_{out} = 2$ (see Figure 5), the velocity profile predicted by the CLB method agree well with the DSMC and IP-DSMC results, and there is very little difference in the pressure deviation. Form Figures 4 and 5 we can observe that, the variation of the pressure deviation distribution from $Kn_{out} = 0.194$ to $Kn_{out} = 0.388$ ($p_{in}/p_{out} = 2$) is decreased with the increase of the rarefaction effects.

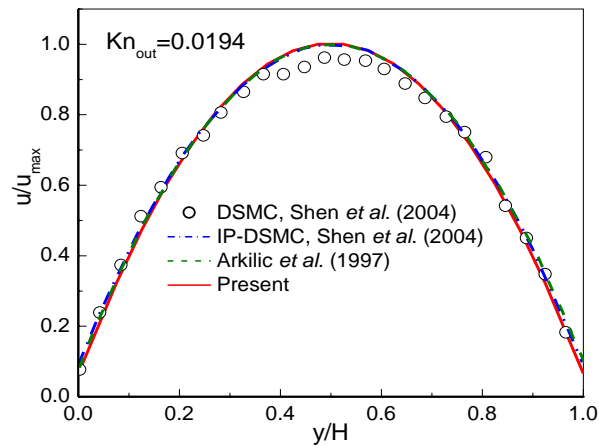

(a)

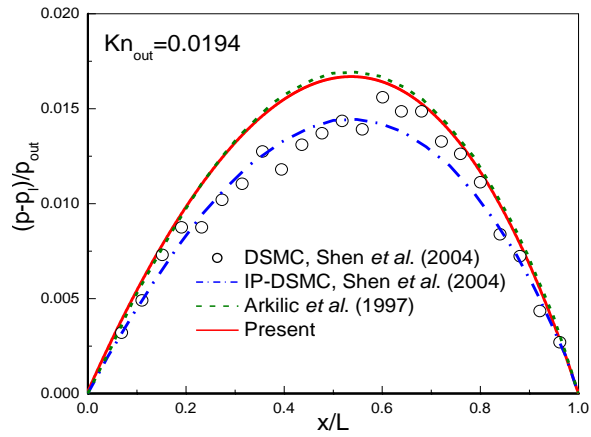

(b)

**Figure 3**. Streamwise velocity at the outlet (a) and pressure deviation along the channel centerline (b) of pressure-driven flow through a long microchannel with $Kn_{out} = 0.0194$ and $p_{in}/p_{out} = 1.4$.

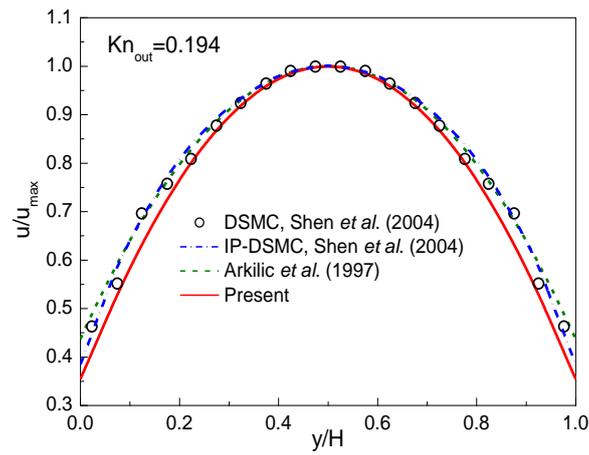

(a)

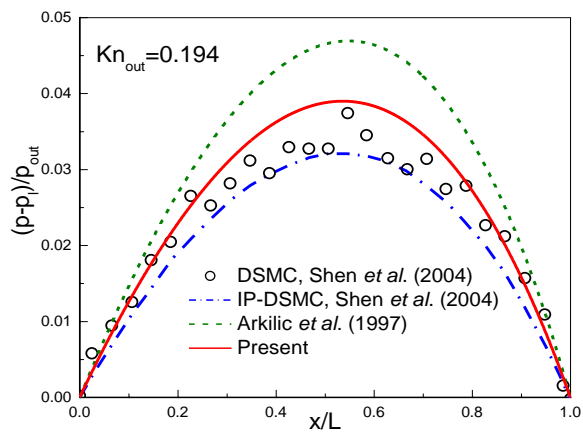

(b)

**Figure 4**. Streamwise velocity at the outlet (a) and pressure deviation along the channel centerline (b) of pressure-driven flow through a long microchannel with $Kn_{out} = 0.194$ and $p_{in}/p_{out} = 2$.

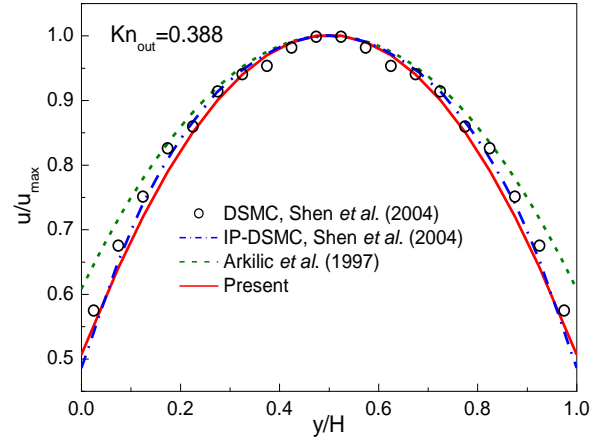

(a)

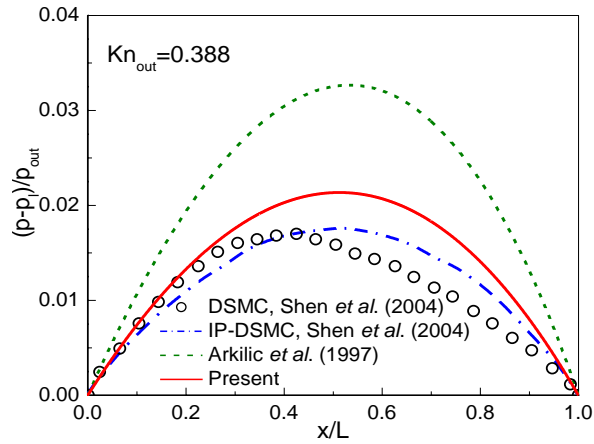

(b)

**Figure 5**. Streamwise velocity at the outlet (a) and pressure deviation along the channel centerline (b) of pressure-driven flow through a long microchannel with $Kn_{out} = 0.388$ and $p_{in}/p_{out} = 2$.

The streamwise velocity ($u_x$) and the spanwise velocity ($u_y$) predicted by the present CLB method for $Kn_{out} = 0.194$ and $p_{in}/p_{out} = 2$ are presented in Figure 6. The streamwise and spanwise velocities are normalized by the maximum streamwise velocity $u_{x,\max}$ at the outlet, i.e., $U = u_x/u_{x,\max}$ and $V = u_y/u_{x,\max}$. Figure 6(a) clearly shows the velocity slip phenomenon at the bottom and top walls, and it can be observed that the slip velocity increases along the microchannel. As shown in Figure 6(b), the magnitude of the spanwise velocity is substantially smaller than that of the streamwise velocity, and the spanwise velocity distribution clearly indicates that the flow migrates from the channel centerline toward the wall as it progresses down the channel. The above observations agree well with those reported in the literature (Guo *et al.*, 2006; Li *et al.*, 2011).

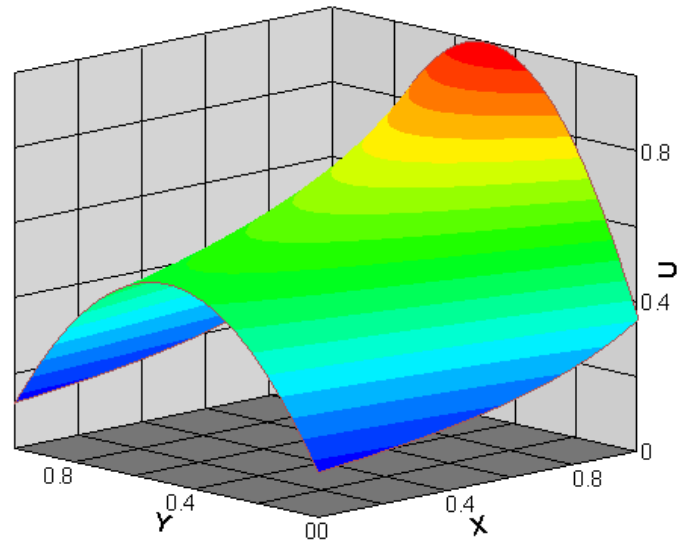

(a)

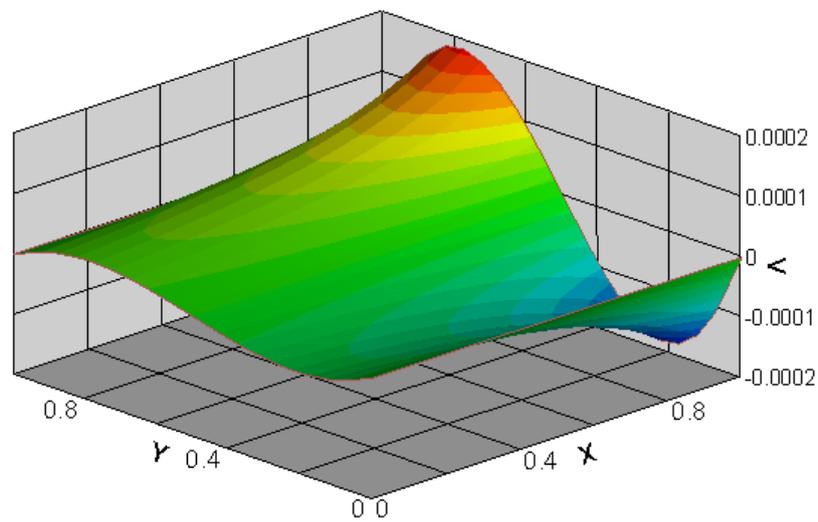

(b)

**Figure 6**. Streamwise (a) and spanwise (b) velocities predicted by the CLB method with $Kn_{out} = 0.194$ and $p_{in}/p_{out} = 2$.

In what follows, the CLB method is applied to study the rarefaction effects on the mass flow rate of pressure-driven flow through a long microchannel. In order to make comparisons with the experimental data for Helium flows by Colin *et al.* (2004) and Maurer *et al.* (2003), the pressure ratio $p_{in}/p_{out}$ and TMAC $\sigma$ are set to be $1.8$ and $0.93$, respectively. The dimensionless mass flow rate is defined as $S = m/m_{ns}$, where $m = \int_0^H (\rho u_x) dy$ is the mass flow rate, and $m_{ns}$ is the corresponding mass flow rate in the absence of rarefaction effects (continuum flow with no-slip boundary condition). In Colin *et al.*'s work, the inverse dimensionless mass flow rate ($1/S$) was plotted against $Kn_{out}$ for Helium flows in a long microchannel up to $Kn_{out} = 0.47$. In Maurer *et al.*'s experiment, the dimensionless mass flow rate ($S$) was plotted against $Kn_{ave} = (Kn_{in} + Kn_{out})/2$ for Helium flows in a long microchannel up to $Kn_{ave} \approx 0.8$. In Figures 7(a) and 7(b), the inverse dimensionless mass flow rate $1/S$ and the dimensionless mass flow rate $S$ are plotted against $Kn_{out}$ and $Kn_{ave}$, respectively. The analytical solutions of Aubert and Colin (2001) and Arkilic *et al.* (1997) are also included in Figure 7 for comparison. From Figure 7(a) we can observe that, Aubert and Colin's second-order slip model and the present CLB method predict nearly the same mass flow rate up to $Kn_{out} = 0.15$. As $Kn_{out}$ increases, Aubert and Colin's analytical solutions gradually deviate from the experimental data, while the present method still gives a satisfactory prediction up to $Kn_{out} = 0.5$. A similar phenomenon can also be observed in Figure 7(b). The above comparisons indicate that by using the Bosanquet-type effective viscosity with the CBBSR boundary scheme, the present CLB method is able to accurately capture the characteristic flow behaviors of pressure-driven gas flow in a long microchannel in the transition flow regime with moderate Knudsen numbers.

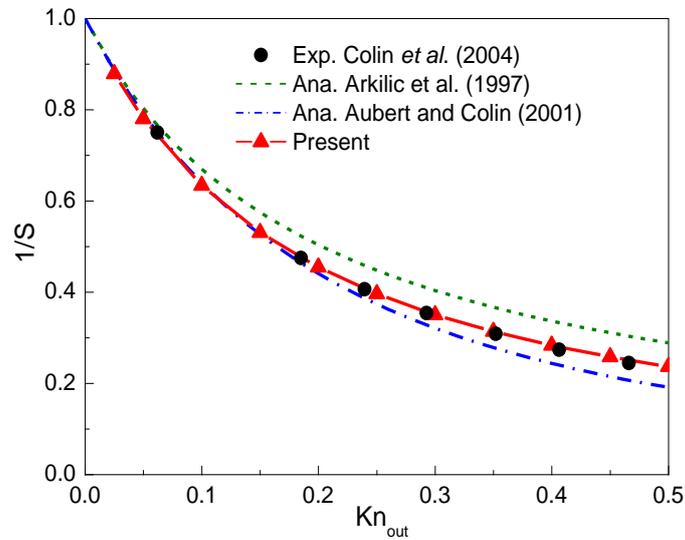

(a)

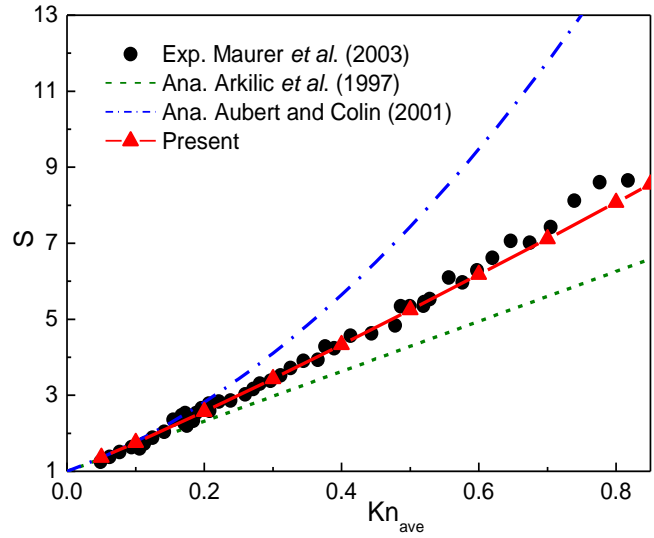

(b)

**Figure 7**. Dimensionless mass flow rate of pressure-driven flow through a long microchannel

$\sigma = 0.93$ and $p_{in}/p_{out} = 1.8$.

## 4. Conclusions

In this paper, a CLB method with the Bosanquet-type effective viscosity has been developed to study microchannel gas flows in the slip and transition flow regimes with a wide range of Knudsen numbers. In order to gain accurate simulations and match the Bosanquet-type effective viscosity, the CBBSR boundary scheme with a modified second-order slip boundary condition is employed. The CLB method is applied to study gas flow in a microchannel with periodic boundary condition and gas flow in a long microchannel with pressure boundary condition from the slip flow regime to the transition flow regime. The predicted results, including the velocity profile, the mass flow rate, and the non-linear pressure deviation distribution along the long microchannel, agree well with the solution of the linearized BE, the DSMC and IP-DSMC results, and the experimental data reported in the literature. The characteristic flow behaviors of microchannel gas flows, such as the Knudsen minimum phenomenon, and the distributions of the streamwise and spanwise velocities, are well captured by the CLB method. In this work, only 2D microchannel gas flows are investigated. The extension of the present CLB method to study 3D microchannel gas flows will be considered in our future studies.

**Acknowledgements**

This work was supported by the Major Program of the National Natural Science Foundation of

China (Grant No. 51590902) and the National Key Basic Research Program of China (973 Program) (2013CB228304).China (Grant No. 51590902) and the National Key Basic Research Program of China (973 Program) (2013CB228304).